# Dimensionality reduction of Poisson's equation with application to particle-in-cell simulations of Hall thrusters

Maryam Reza [*1], Farbod Faraji[*2], Aaron Knoll[*3]

[*]Plasma Propulsion Laboratory, Department of Aeronautics, Imperial College London, Exhibition Road, London, SW7 2AZ, United Kingdom

[1] Corresponding Author (m.reza20@imperial.ac.uk); [2] f.faraji20@imperial.ac.uk; [3] a.knoll@imperial.ac.uk

**Abstract**

In this article, we introduce a novel dimensionality reduction formulation for the Poisson's equation in the Vlasov-Poisson system that yields a reduced-order particle-in-cell scheme. This scheme allows a remarkable reduction in the computational cost of self-consistent kinetic simulations of Hall thrusters. The formulation of the dimensionality reduction approach, together with its verification for a general Poisson's problem, is presented. Moreover, we show the results of several "quasi-2D" axial-azimuthal simulations we performed with the conditions of a well-established 2D3V reference case. Comparison between the results of the quasi-2D simulations and the reference case revealed that our method is able to resolve accurately the axial distributions of the intensive plasma parameters. In addition, the characteristics of the azimuthal instabilities observed from the quasi-2D simulation are shown to be closely in line with those reported in the literature. Accordingly, the reduced-order PIC scheme is verified as a viable, low-computational-cost alternative to the traditional multi-dimensional particle-in-cell simulations of Hall thrusters.

---

Hall thrusters, first conceived in the early 1960s, have become one of the leading technologies for spacecraft Electric Propulsion. As the power and size requirements of these devices increase, as they have in recent years, so too does the cost, time, and associated risks for their development [1]. Self-consistent kinetic simulations have a sufficient level of fidelity to be used as predictive numerical tools, supporting the development and qualification of these new technologies. However, these simulations are hindered significantly by the computational constraints of current High Performance Computing facilities [2].

In a previous publication [3], we introduced a novel approach to enhance the predictive capability of one-dimensional axial particle-in-cell (PIC) simulations of Hall thrusters to self-consistently resolve the electrons' cross-field mobility induced by the azimuthal instabilities. Our proposed approach essentially involved solving in parallel 1D axial and azimuthal PIC simulations that "share" the same macroparticles. We demonstrated for several test cases that such a "pseudo-2D" PIC simulation provides predictions that compare very closely with those from full-2D axial-azimuthal simulations, but at a fraction of the computational cost.

In this article, we present a dimensionality-reduction formulation to decompose the multi-dimensional Poisson's equation into a coupled system of 1D Ordinary Differential Equations. The incorporation of this formulation in a PIC code yields a new scheme, which we term "reduced-order" PIC. The underpinning dimensionality-reduction approach allows us to generalize the concept of the pseudo-2D PIC simulation based on a mathematically rigorous treatment of the decomposition of the Poisson's equation. The general reduced-order PIC scheme still offers a significant reduction in the required total number of computational cells and, thus, the necessary total number of macroparticles.

However, before looking into the computational advantage of the reduced-order PIC scheme, we focus on the mathematical formulation of the dimensionality-reduction approach.

Without losing generality, we focus from this point forward on a two-dimensional domain, and thus, the 2D Poisson's equation. To introduce the basic concepts behind the dimensionality-reduction approach, we first refer to Figure 1 in which a 2D domain with the extents $L_x$ and $L_y$ along the x- and y-direction, respectively, is represented by two perpendicular computation grids along x and y. The computational cells along the x-direction have the width $\Delta x$, determined according to the physics of the problem, and are extended over the entire $L_y$ extent of the domain. Similarly, the computational cells along y have the width $\Delta y$ and a length equal to $L_x$. We term this representation of the domain the "single-region" decomposition, since we have a single set of perpendicular grids representing the entire domain.

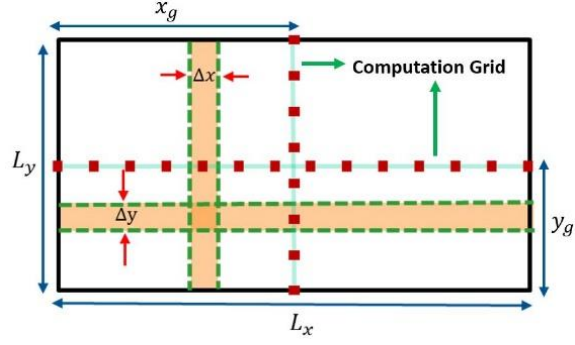

Figure 1: Schematic of the computational domain and the defined "Region" for the single-region implementation

We define two potential functions $\phi^x$ and $\phi^y$, which are, by definition, only a function of the x- and y-coordinate, respectively, within a region. We further assume that the 2D potential function $\phi(x,y)$ can be approximated as the superimposition of these two potential functions, i.e.,

$$\phi(x,y) = \phi^x(x) + \phi^y(y). \tag{Eq. 1}$$

Substituting Eq. 1 in the 2D Poisson's equation

$$\nabla^2 \phi(x,y) = \frac{\partial^2 \phi(x,y)}{\partial x^2} + \frac{\partial^2 \phi(x,y)}{\partial y^2} = -\frac{\rho(x,y)}{\epsilon_0}, \tag{Eq. 2}$$

and integrating once along the horizontal (x) and once along the vertical (y) computation grid yields a coupled system of 1D Ordinary Differential Equations for $\phi^x$ and $\phi^y$, respectively,

$$\frac{\partial^2 \phi^x(x)}{\partial x^2} L_y + \frac{\partial \phi^y}{\partial y}\Big|_{y_g+\frac{L_y}{2}} - \frac{\partial \phi^y}{\partial y}\Big|_{y_g-\frac{L_y}{2}} = -\frac{1}{\epsilon_0} \int_{y_g-\frac{L_y}{2}}^{y_g+\frac{L_y}{2}} \rho(x,y) dy, \tag{Eq. 3}$$

$$\frac{\partial^2 \phi^y(y)}{\partial y^2} L_x + \frac{\partial \phi^x}{\partial x}\Big|_{x_g+\frac{L_x}{2}} - \frac{\partial \phi^x}{\partial x}\Big|_{x_g-\frac{L_x}{2}} = -\frac{1}{\epsilon_0} \int_{x_g-\frac{L_x}{2}}^{x_g+\frac{L_x}{2}} \rho(x,y) dx. \tag{Eq. 4}$$

In the above equations, $\epsilon_0$ is the permittivity of free space, and $x_g$ and $y_g$ are the locations of the x and y computation grid, respectively, with respect to the origin. $\rho(x,y)$ is the 2D charge density distribution.

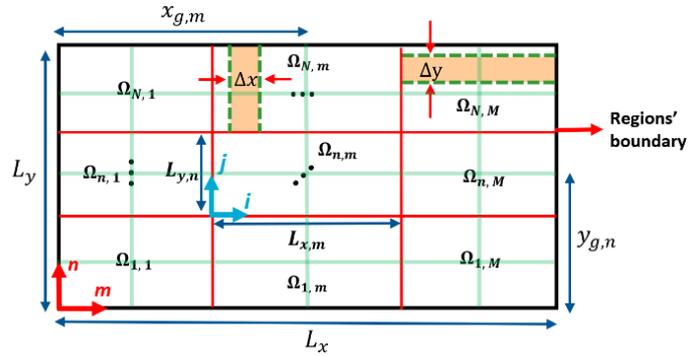

Figure 2: Schematic of the "Multi-Region" domain decomposition for the dimensionality reduction of the Poisson's equation

Now, referring to Figure 2, the described single-region decomposition is extended to "multi-region" where the 2D domain is decomposed into a grid of $N \times M$ regions, denoted by $\Omega_{n,m}$. The regions' boundary is defined such that it is equidistant from the two computational nodes adjacent to it. Within each region, there is a set of perpendicular grids along which the 1D potential functions $\phi^x_{n,m}$ and $\phi^y_{n,m}$ are solved. Moreover, the 2D potential function is now approximated as

$$\phi(x,y) = \phi_{n,m}^x + \phi_{n,m}^y, \quad \text{with} \quad x,y \in \Omega_{n,m} \tag{Eq. 5}$$

Accordingly, Eqs. 3 and 4 are written as in Eqs 6 and 7 for the multi-region decomposition, yielding the general formulation of the dimensionality-reduction approach:

$$\begin{aligned}
&\left(\frac{\partial^2 \phi_{n,m}^x}{\partial x^2}\bigg|_{x=x_i}\right) L_{y,n} + \left(\int_{y_{g,n}-\frac{L_{y,n}}{2}}^{y_{g,n}+\frac{L_{y,n}}{2}} \left(\frac{\partial \phi_{n,m}^y}{\partial x}\bigg|_{x=x_i+\frac{\Delta x}{2}}\right) dy - \int_{y_{g,n}-\frac{L_{y,n}}{2}}^{y_{g,n}+\frac{L_{y,n}}{2}} \left(\frac{\partial \phi_{n,m}^y}{\partial x}\bigg|_{x=x_i-\frac{\Delta x}{2}}\right) dy\right)\frac{1}{\Delta x} \\
&+ \left(\frac{\partial \phi_{n,m}^y}{\partial y}\bigg|_{y=y_{g,n}+\frac{L_{y,n}}{2}} + \frac{\partial \phi_{n,m}^x}{\partial y}\bigg|_{y=y_{g,n}+\frac{L_{y,n}}{2}}\right) \\
&- \left(\frac{\partial \phi_{n,m}^y}{\partial y}\bigg|_{y=y_{g,n}-\frac{L_{y,n}}{2}} + \frac{\partial \phi_{n,m}^x}{\partial y}\bigg|_{y=y_{g,n}-\frac{L_{y,n}}{2}}\right) = -\frac{1}{\epsilon_0}\int_{y_{g,n}-\frac{L_{y,n}}{2}}^{y_{g,n}+\frac{L_{y,n}}{2}} \rho(x,y)dy,
\end{aligned} \tag{Eq. 6}$$

$$\begin{aligned}
&\left(\frac{\partial^2 \phi_{n,m}^y}{\partial y^2}\bigg|_{y=y_j}\right) L_{x,m} + \left(\int_{x_{g,m}-\frac{L_{x,m}}{2}}^{x_{g,m}+\frac{L_{x,m}}{2}} \left(\frac{\partial \phi_{n,m}^x}{\partial y}\bigg|_{y=y_j+\frac{\Delta y}{2}}\right) dx - \int_{x_{g,m}-\frac{L_{x,m}}{2}}^{x_{g,m}+\frac{L_{x,m}}{2}} \left(\frac{\partial \phi_{n,m}^x}{\partial y}\bigg|_{y=y_j-\frac{\Delta y}{2}}\right) dx\right)\frac{1}{\Delta y} \\
&+ \left(\frac{\partial \phi_{n,m}^x}{\partial x}\bigg|_{x=x_{g,m}+\frac{L_{x,m}}{2}} + \frac{\partial \phi_{n,m}^y}{\partial x}\bigg|_{x=x_{g,m}+\frac{L_{x,m}}{2}}\right) \\
&- \left(\frac{\partial \phi_{n,m}^x}{\partial x}\bigg|_{x=x_{g,m}-\frac{L_{x,m}}{2}} + \frac{\partial \phi_{n,m}^y}{\partial x}\bigg|_{x=x_{g,m}-\frac{L_{x,m}}{2}}\right) = -\frac{1}{\epsilon_0}\int_{x_{g,m}-\frac{L_{x,m}}{2}}^{x_{g,m}+\frac{L_{x,m}}{2}} \rho(x,y)dx.
\end{aligned} \tag{Eq. 7}$$

In Eqs. 6 and 7, $L_{x,m}$ and $L_{y,n}$ are the horizontal and vertical extents of each region $\Omega_{n,m}$, $x_i$ is the x coordinate of each computational node $i$ along a horizontal grid and $y_j$ is similarly the y coordinate of each node $j$ along a vertical grid.

Having solved the coupled system of equations given by Eqs. 6 and 7 with appropriate boundary conditions, to be described shortly, the electric field is determined within each region using $\vec{E} = -\vec{\nabla}\phi(x,y) = -\vec{\nabla}\left(\phi_{n,m}^x(x) + \phi_{n,m}^y(y)\right)$.

In Eqs 6 and 7, it is important to note the appearance of the so-called "cross-derivative" terms, $\frac{\partial \phi_{n,m}^x}{\partial y}$ and $\frac{\partial \phi_{n,m}^y}{\partial x}$, that take into account the variations in $\phi_{n,m}^x$ and $\phi_{n,m}^y$ when transiting from one region to another along the y- and x-direction, respectively.

Moreover, recalling that, inside each region $\Omega_{n,m}$, $\phi_{n,m}^x$ is, by definition, constant along y, and $\phi_{n,m}^y$ is likewise constant along x, the integral terms within the second parentheses on the left-hand-side of Eqs. 6 and 7 are zero for the computational cells that reside entirely inside each region.

In other words, the integrand terms $\frac{\partial \phi_{n,m}^y}{\partial x}\big|_{x=x_i+\frac{\Delta x}{2}}$ and $\frac{\partial \phi_{n,m}^y}{\partial x}\big|_{x=x_i-\frac{\Delta x}{2}}$ in Eq. 6 and, similarly, $\frac{\partial \phi_{n,m}^x}{\partial y}\big|_{y=y_j+\frac{\Delta y}{2}}$ and $\frac{\partial \phi_{n,m}^x}{\partial y}\big|_{y=y_j-\frac{\Delta y}{2}}$ in Eq. 7 are *non-zero* only when one side of a computational cell falls on a region's boundary.

The boundary conditions accompanying Eqs. 6 and 7 are of the following general form

$$\begin{cases}
\phi_{n,m}^x(x_b) + \phi_{n,m}^{y0} = \dfrac{1}{L_{y,n}}\displaystyle\int_{y_{g,n}-\frac{L_{y,n}}{2}}^{y_{g,n}+\frac{L_{y,n}}{2}} \phi(x_b,y)\,dy, \\[2mm]
\phi_{n,m}^y(y_b) + \phi_{n,m}^{x0} = \dfrac{1}{L_{x,m}}\displaystyle\int_{x_{g,m}-\frac{L_{x,m}}{2}}^{x_{g,m}+\frac{L_{x,m}}{2}} \phi(x,y_b)\,dx,
\end{cases} \tag{Eq. 8}$$

in which, $x_b$ and $y_b$ are the locations of the domain boundary along the x and y-direction, respectively. In addition,

$$\phi_{n,m}^{y0} = \frac{1}{L_{y,n}}\int_{y_{g,n}-\frac{L_{y,n}}{2}}^{y_{g,n}+\frac{L_{y,n}}{2}} \phi_{n,m}^y(y)dy, \quad \text{and,} \tag{Eq. 9}$$

$$\phi_{n,m}^{x0} = \frac{1}{L_{x,m}} \int_{x_{g,m}-\frac{L_{x,m}}{2}}^{x_{g,m}+\frac{L_{x,m}}{2}} \phi_{n,m}^{x}(x)dx. \tag{Eq. 10}$$

The formulation presented above for the dimensionality-reduction approach provides a generalizable "quasi-multi-dimensional" approximation of the Poisson's problem which is included in the PIC scheme to yield the reduced-order PIC.

Following the discussion of the dimensionality-reduction formulation, we demonstrate below for a 2D case the computational advantage that the reduced-order PIC has over a traditional two-dimensional simulation. Table 1 provides a comparison between a full-2D and a quasi-2D PIC scheme in terms of two simulation parameters, number of cells ($N_{cells}$) and the total number of macroparticles at the beginning of the simulation ($N_{p,init}$).

|  | $N_{cells}$ | $N_{PPC}$ | $N_{p,init}$ |
|---|---|---|---|
| **Full-2D** | $N_i \times N_j = 128000$ | 150 | $1.92 \times 10^7$ |
| **Quasi-2D** (N × M regions) | $\max(N \times N_i, M \times N_j)$ | 150 | $150 \times \max(N \times N_i, M \times N_j)$ |
| **Quasi-2D** (Single region) | $\max(N_i, N_j) = 500$ | 150 | $7.5 \times 10^4$ |

Table 1: Comparison between a full-2D and quasi-2D PIC simulation in terms of the grid size ($N_{cells}$) and the total number of macroparticles at the simulation start ($N_{p,init}$); $N_{PPC}$ is the number of macroparticles per cell.

For the comparison in Table 1, we have considered a domain with $N_i = 500$ cells along the x-direction and $N_j = 256$ cells along the y-direction. These are the same number of computational cells used for the full-2D axial-azimuthal benchmark in [4], with which we later compare the results of our reduced-order simulations.

Looking at the last column of Table 1, the single-region quasi-2D PIC results in 256 times reduction in the computational cost. However, it is important to note that decomposing the domain even into multiple regions (e.g., $M = N = 10$), to account for possible effects of the dominant gradients in the plasma properties, still translates into over an order of magnitude computational gain.

This gain is reflected directly in the necessary computational resources. Indeed, for the benchmark simulation case of [4], the conventional 2D kinetic codes of the participating groups required about 2.5 to 11 days when using about 100 to 360 CPU cores. Whereas, our simulations of the same case took less than 1 day using only 16 CPU cores.

The verification of our approach has been performed in two steps: first, the dimensionality-reduction formulation was verified for a general test problem. Second, we set up quasi-2D PIC simulations with the conditions similar to those in the well-established full-2D axial-azimuthal benchmark of Ref. [4] and compared the axial distribution of plasma parameters in addition to the characteristics of the resolved azimuthal waves. The so-called "single-region" ($M = 1, N = 1$), "double-region" ($M = 2, N = 1$) and "triple-region" ($M = 3, N = 1$) reduced-order PIC simulations were carried out.

We first focus on the results of the verification of the dimensionality-reduction formulation. The chosen test case is a 2D Poisson's problem [8], defined as $\phi_{xx} + \phi_{yy} = x^2 + y^2$ for the variable $\phi(x, y)$ with the boundary conditions $\phi(x, 0) = 0$, $\phi(x, L_y) = \frac{1}{2}x^2$, $\phi(0, y) = \sin(\pi y)$, $\phi(L_x, y) = \exp(\pi L_x)\sin(\pi y) + \frac{1}{2}y^2$, over a domain with the extents $L_x = L_y = 1$. This problem has the analytical solution $\phi(x, y) = \exp(\pi x)\sin(\pi y) + \frac{1}{2}(xy)^2$.

Plots (a)-(c) in Figure 3 show the numerical approximations of the 2D analytical solution (plot (d)) for various number of regions $N$ and $M$, with the number of cells $N_i = N_j = 200$. In addition, Figure 4 illustrates the variation of the L2 error in the approximate solutions with respect to the analytical one when increasing $N$ and $M$ simultaneously.

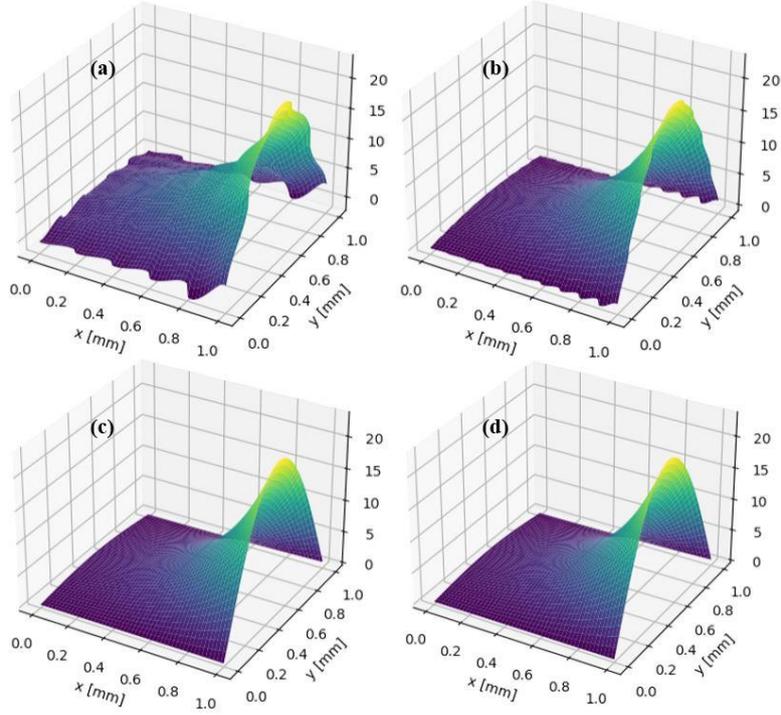

Figure 3: Dimensionality-reduction approximations of the 2D Poisson's problem for various number of regions; (a) $M = N = 5$, (b) $M = N = 20$, (c) $M = N = 50$, and (d) the analytical solution.

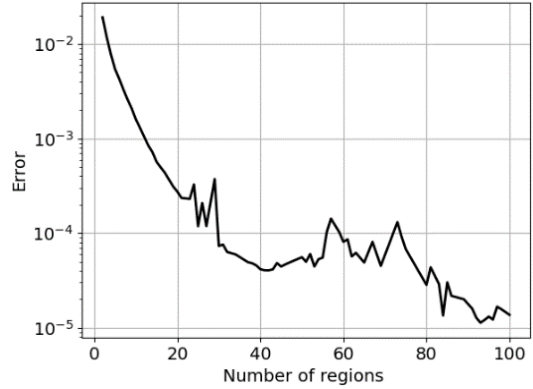

Figure 4: Variation of the L2 error between the exact and approximate solutions from the dimensionality-reduction approach when increasing $M$ and $N$. The y-axis is in the logarithmic scale.

Looking at the plots in Figure 3, it is clear that, despite the complex nature of the analytical solution, the dimensionality-reduction approach is able to capture all of the solution's main features (e.g., the symmetry in the y-direction and the convex curvature along x) even with a low number of regions of $M = N = 5$ (Figure 3(a)).

Moreover, increasing the number of regions to the $M = N = 50$, the approach is able to very closely approximate the analytical solution. This can be better understood from Figure 4, in which the L2 error is seen to fall very quickly when increasing the number of regions, approaching error values in the same order of a full-2D Poisson solver. Concerning the rapid decrease in the error, it is worth noting that this is indeed a desirable behavior considering that, in a PIC simulation, we are interested in implementing the dimensionality-reduction approach at the limit of low number of regions.

Furthermore, from $M = N = 5$, the L2 error is always below 1%, reaching values in the order of $10^{-3}$% for the $M = N = 100$, thus, replicating the exact solution in the limit. The reason why the number of regions in this test case (a 2D $200 \times 200$ grid) is not increased beyond 100 is that our dimensionality-reduction approach requires that at least two nodes fall within each region.

Following the above verification, we now compare the predictions of our axial-azimuthal quasi-2D PIC simulations against the benchmark's results [4]. The baseline PIC code, in which the dimensionality-reduction approach is implemented, alongside some of the code's benchmarking results, are presented in Ref. [3].

The simulation setup is similar to the one presented in Ref. [4]. The simulation domain is a Cartesian $(x-y)$ plane with $x$ along the axial coordinate and $y$ along the azimuth. The domain's extents, cell size, time step, total simulation time and the initial plasma and boundary conditions are those reported in Table 1 of Ref. [4]. The simulations were always started with 300 macroparticles per cell, with the total initial number of macroparticles being calculated according to the relation $300 \times \max(N \times N_i, M \times N_j)$ for the different quasi-2D simulations performed.

Figure 5 presents the comparison with the benchmark of the time-averaged plasma properties, electric field ($E_x$), electron temperature ($T_e$), and ion density ($n_i$). Profiles of these parameters from quasi-2D simulations are compared with the mean value of the corresponding profiles from different PIC codes reported in Ref. [4].

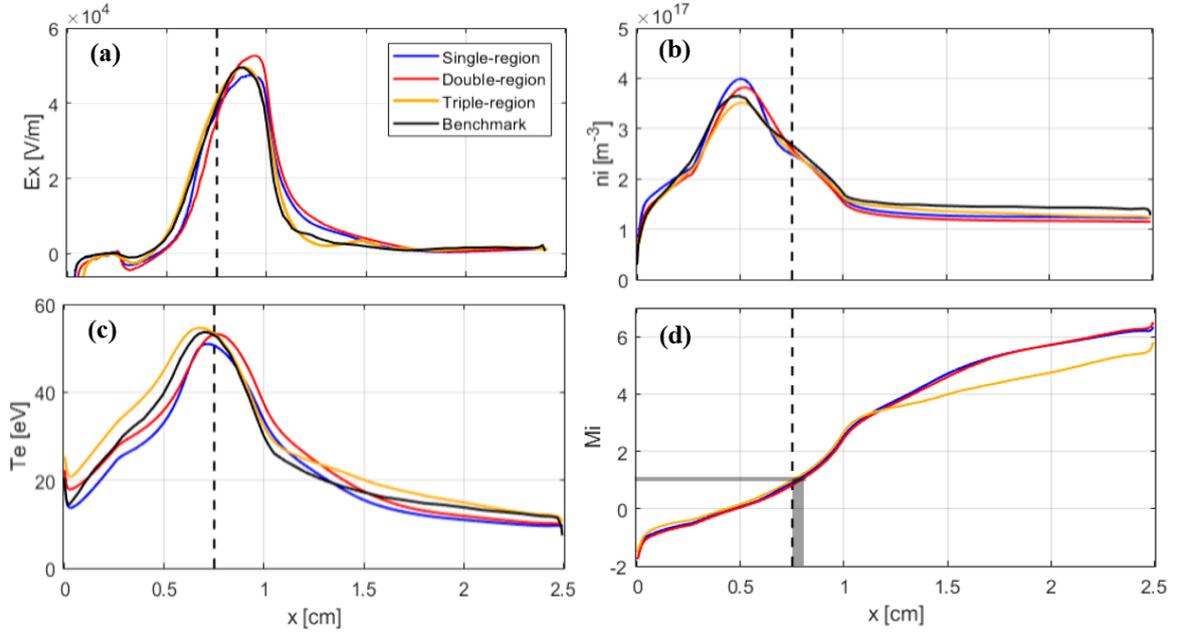

Figure 5: Axial profiles of plasma properties averaged over 16-20 μs: (a) $E_x$, (b) $n_i$, (c) $T_e$. In (d), the axial distribution of ion Mach number ($M_i$) is shown. Black dashed lines show the location of peak magnetic field.

From plots (a)-(c) in Figure 5, it is observed that all quasi-2D simulations provide a remarkably similar prediction of the time-averaged plasma properties distribution compared with the results of the full-2D benchmark simulation.

In Figure 5(d), we have additionally shown the axial distribution of the ion Mach number, with the grey-shaded area highlighting the axial positions where the ion sonic point occurs in the quasi-2D simulations. It is observed that the sonic point almost corresponds to the location of the peak magnetic field intensity. This observation is in line with that reported in Refs. [4][6].

Furthermore, it is noted that, in the triple-region simulation, the boundary between the first and second vertical region ($M = 1$ and $M = 2$) was set at the location of the maximum magnetic field intensity, corresponding in this case exactly to the ion sonic point (Figure 5(d)). This is because recent evidence [4][7] suggests that the ion sonic point corresponds to an axial transition location in the characteristics of the azimuthal instabilities. Accordingly, the analyses in the remainder of this article on the waves' characteristics are done using the triple-region results. Nonetheless, it is interesting from the plots in Figure 5 that, at least concerning the macroscopic plasma properties, the quasi-2D simulations' prediction show a negligible sensitivity to the location of the boundary between the regions.

*Analysis of the azimuthal instabilities*
A 2D Fast Fourier Transform (FFT) was applied to the azimuthal electric field signals from the triple-region simulation (Figure 6(a)-(c)), after reaching steady state, to obtain the dispersion map in the $(k_y - \omega_R)$ plane, with $k_y$ being the azimuthal wave number and $\omega_R$ being the real frequency component. Figure 7(a)-(c) show the dispersion plots for the waves captured in regions I to III of the triple-region simulation. The theoretical dispersion relation of the ion acoustic waves in the ions' reference frame (Eq. 11) [8] is also plotted in all maps.

$$\omega_R \approx \pm \frac{k_y C_s}{\sqrt{1 + k_y^2 \lambda_D^2}}. \qquad \text{(Eq. 11)}$$

In Eq. 11, $C_s$ is the ion sound speed, and $\lambda_D$ is the Debye length. In the plots of Figure 7, the yellow lines are the local dispersion relation plots, calculated using the axially averaged plasma properties in each region, whereas the green lines are the ion acoustic dispersion relation corresponding to the plasma properties in Region I.

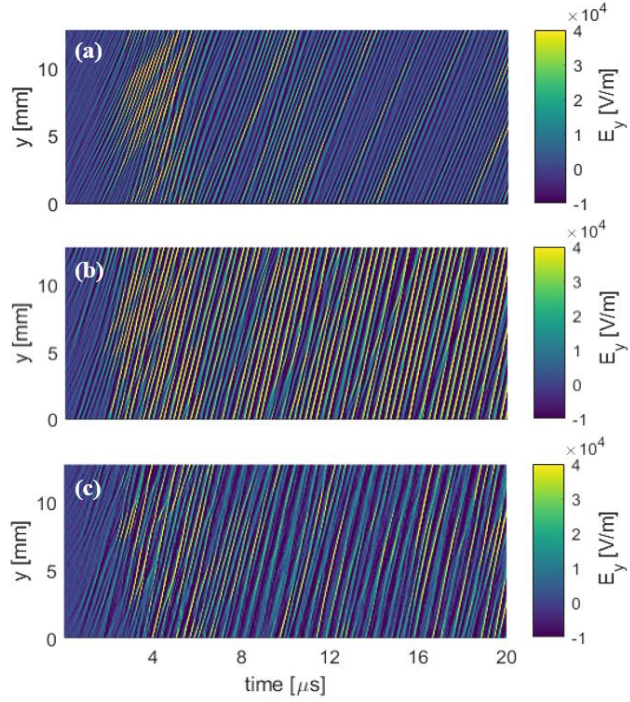

Figure 6: Spatiotemporal maps of the azimuthal electric field fluctuations from the triple-region quasi-2D simulation; (a) Region I, (b) Region II, (c) Region III.

Referring, first, to Figure 6, the transition in the waves' characteristics from Region I (Figure 6(a)) to II (Figure 6(b)) is clear, which is consistent with the full-2D simulation results [4][7]. In addition, it is observed that the azimuthal waves in Region II are of higher amplitude compared to the waves in Region III (Figure 6(c)), possibly due to stronger wave-particle interactions in this region as a result of a higher plasma density than that in Region III.

Second, in Figure 7, it is observed that the dispersion characteristics in Region I (Figure 7(a)) are in very close agreement with the dispersion relation of the ion acoustic waves (Eq. 11). This can be justified by noting that, when the simulation arrives at steady state, the excited waves are in the pseudo-saturated state, which is shown in the literature to be accompanied by a transition to the ion acoustic waves [9]. Furthermore, it is seen that, going from Region I to II (Figure 7(b)) and III (Figure 7(c)), the waves' dispersion plots increasingly deviate from the *local* ion acoustic dispersion relation (yellow line). However, they are closely consistent with the ion acoustic dispersion relation of Region I (green line). This demonstrates that the quasi-2D simulation is capturing the downstream convection by the ion beam of the waves excited in Region I, a phenomenon that has been reported as well in Refs. [7][10].

The remarkable consistency between the observations above from the quasi-2D and full-2D simulations serves as an additional proof that the reduced-order PIC is not only capable of correctly predicting the time-averaged plasma properties, but it also provides an accurate picture of the involved physics, including the multi-dimensional phenomena.

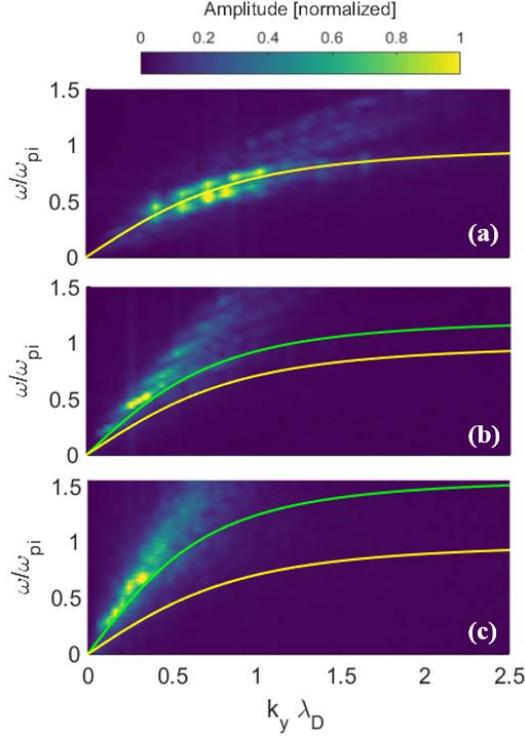

Figure 7: 2D FFT plots of the azimuthal electric field fluctuations from the triple-region simulation; (a) Region I, (b) Region II, (c) Region III. Yellow lines are the local dispersion relations, and the green lines are the dispersion relation of the ion acoustic waves in Region I.

*Contribution to transport of different terms in the electron azimuthal momentum equation*

Finally, we examine the contribution of various terms in the electron azimuthal momentum equation to the cross-field transport. The approach pursued here is similar to that in Refs. [11] and [3]. The electron momentum equation along the azimuthal direction $y$ can be written as

$$-qn_e v_{e,x} B = \partial_t(mn_e v_{e,y}) + \partial_x(mn_e v_{e,x} v_{e,y}) + \partial_x(\Pi_{e,xy}) - qn_e E_y,$$ (Eq. 12)

where, $q$ is the elementary charge, $n_e$ is the electron number density, $v_{e,x}$ and $v_{e,y}$ are the electron axial and azimuthal drift velocity, $B$ is the magnetic field intensity, $m$ is the electron mass and $E_y$ is the azimuthal electric field. In Eq. 12, the left-hand-side term is called the magnetic force ($F_B$) whereas, on the right-hand-side (RHS), the first term is the temporal inertia force ($F_t$), the second term is the convective inertia force ($F_I$), the third force term corresponds to the viscous effects ($F_{\Pi_{xy}}$), and the last term is the electric force term ($F_E$). We discussed in Ref. [3] that the term $F_E$ actually represents the contribution of the azimuthal instabilities to cross-field transport in our simulations. The derivations of the force terms in Eq. 12 are provided in Ref. [3].

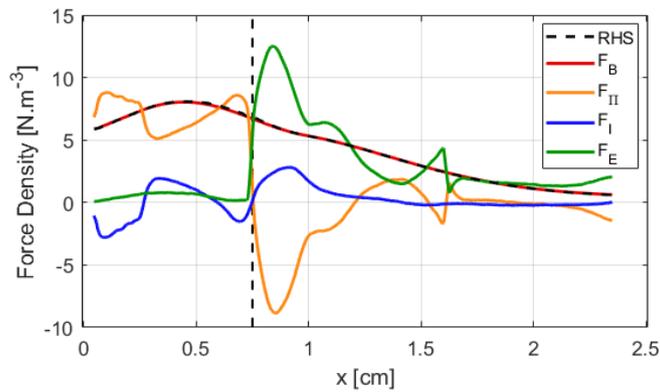

Figure 8: Axial distribution of the terms in Eq. 12 for the triple-region simulation. The force terms are averaged over 4 $\mu s$. $F_t$ contribution was almost zero, and it is thus not displayed.

Figure 8 depicts the axial profiles of the magnetic force term and each of the terms on the RHS of Eq. 12 from the triple-region simulation. The dashed black curve corresponds to the sum of all RHS terms, which is perfectly equal to the magnetic force term ($F_B$), demonstrating the satisfaction of Vlasov's equation. Thus, it is concluded that no notable numerical effects have influenced the electron's cross-field transport in our simulation.

It is additionally pointed out that the axial distributions of the momentum terms in Figure 8 and their relative magnitudes were found to be similar to those in Ref. [7].

To conclude, the verification results of the Poisson's dimensionality-reduction approach and the similarity of our quasi-2D PIC results to those of the benchmark demonstrate the viability of the reduced-order PIC scheme as a computationally efficient, high-fidelity alternative to the current multi-dimensional PIC schemes.

Moreover, it is emphasized that the results presented here demonstrate that capturing an average effect of the azimuthal instabilities on electrons' mobility, even with a single-region approximation, can be sufficient to accurately resolve the axial distribution of the macroscopic plasma properties. Nevertheless, increasing the number of regions can allow the reduced-order PIC scheme to provide a more accurate picture of the underlying physical mechanisms since the simulation is expected to practically converge to a fully multi-dimensional one. As a result, we are currently performing high number-of-region reduced-order simulations to further verify this assertion.

## Acknowledgments

The present research is carried out within the framework of the project "Advanced Space Propulsion for Innovative Realization of space Exploration (ASPIRE)". ASPIRE has received funding from the European Union's Horizon 2020 Research and Innovation Programme under the Grant Agreement No. 101004366. The views expressed herein can in no way be taken as to reflect an official opinion of the Commission of the European Union.